\documentclass[aps,pra,reprint,superscriptaddress]{revtex4-2}

\usepackage{amsmath}
\usepackage{amsfonts}
\usepackage{amssymb}
\usepackage{multirow}
\usepackage[colorlinks=true, allcolors=blue]{hyperref}
\usepackage{orcidlink}

\begin{document}
\title{Partial solvability induced by dark states in a box trap with decentered two-body interaction}

\author{H.~Abedi \orcidlink{0009-0001-8543-1082}}
\affiliation{Department of Physics, University of Tehran, Tehran 14395-547, Iran}

\author{N.~L.~Harshman \orcidlink{0000-0003-2655-3327}}
\affiliation{Physics Department, American University, Washington DC 20016, USA}

\author{P.~Schmelcher \orcidlink{0000-0002-2637-0937}}
\affiliation{Center for Optical Quantum Technologies, Department of Physics, University of Hamburg, Luruper Chaussee 149, 22761 Hamburg, Germany}

\date{\today}

\begin{abstract}
We consider a generalization of the two-body contact interaction for nonrelativistic particles confined to a one-dimensional box,
in which the interaction is decentered, i.e., the particles interact only when they are separated by a distance $c$.
In contrast to the harmonically trapped system, this model is nonintegrable.
Despite this, we demonstrate that the system exhibits partial solvability due to the presence of dark states,
i.e., bosonic or fermionic states unaffected by the interaction.
These states form exactly solvable subspaces embedded within an interacting spectrum.
We characterize the stationary properties of the system,
identify the conditions for the appearance of dark states,
and show how they structure the spectrum and delineate interacting and noninteracting sectors.
\end{abstract}

\maketitle

\section{Introduction and setup}
Exact solvability in interacting quantum systems has long been appreciated for its role in advancing our understanding of quantum systems in general \cite{sutherland2004beautiful}.
In essence, exact solvability indicates that physical quantities can be expressed via a set of algebraic equations.
Since Bethe's pioneering work, which obtained the exact solution of the Heisenberg magnetic spin chain through what is now known as the Bethe ansatz \cite{betheZurTheorieMetalle1931},
the method has been developed and applied to obtain exact solutions in a wide range of one-dimensional (1D) interacting quantum systems \cite{Gaudin_2014, franchini2017introduction}.
As a particularly tractable case, two-body quantum systems serve as paradigmatic models that provide insight into many-body interacting physics \cite{buschTwoColdAtoms1998}.

One important class of such systems is those with two-body contact interactions,
which have proven to be particularly useful for describing the physics of ultracold atomic gases confined to (effectively) 1D geometries where the relevant length scales are much larger than the effective range due to the underlying atomic interaction potential \cite{blochManyBodyPhysicsUltracold2008, olshaniiAtomicScatteringPresence1998}.
Unlike in higher dimensions \cite{BRAATEN2006259}, zero-range two-body interactions in 1D leads to a well-defined many-body problem,
without the need for regularization \cite{buschTwoColdAtoms1998, Farrell_2010}.
This zero-range interaction underlies many prominent exactly solvable models, such as the Lieb-Liniger model for the Bose gas \cite{PhysRev.130.1605}, and the Yang-Gaudin model for the Fermi gas \cite{PhysRevLett.19.1312, GAUDIN196755}.
These models have since inspired extensive theoretical developments \cite{Batchelor_2005, Oelkers_2006},
and their exact predictions have been confirmed by ultracold atomic gas experiments across various interaction regimes \cite{RevModPhys.85.1633, RevModPhys.83.1405}. 

In the few-body regime, trapped systems with contact interaction in 1D are of particular interest,
as ultracold atomic platforms provide unprecedented experimental control to explore a wide range of few-body quantum effects \cite{mistakidisFewbodyBoseGases2023, Blume_2012, Sowiński_2019}.
A remarkable result is the Bose-Fermi mapping \cite{girardeau1960relationship},
which shows that in the strongly interacting limit, bosonic wavefunctions,
for arbitrary trapping potentials, can be mapped one-to-one onto those of noninteracting fermions,
with an identical energy spectrum.

However, zero-range contact interactions have intrinsic limitations from a physical perspective.
One key limitation is their inability to describe the physics of odd-wave interactions.
Due to the symmetry $\delta(x_2 - x_1) = \delta(x_1 - x_2)$,
the contact interaction acts exclusively in the even-wave channel \cite{mistakidisFewbodyBoseGases2023}.
Consequently, single-component fermions are \emph{dark} to zero-range contact interactions,
which correspond to the s-wave scattering limit at ultracold temperatures:
their antisymmetric spatial wavefunctions vanish at the two-body coincidence points, where the delta potential acts,
thus spin-polarized fermions remain insensitive to the interaction. 

\begin{figure}[t]
	        \centering
	        \includegraphics[width= 0.55 \linewidth]{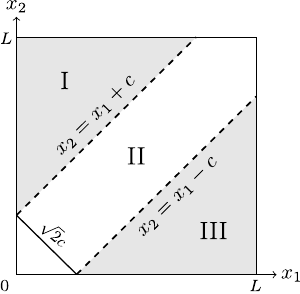}
	        \caption{Configuration space for two particles confined in a box trap of length $L$.
	        The interaction potential has support on the two lines $x_2 - x_1 = \pm c$ 
	        (dashed lines), which divides the space into three regions.}
	    \label{Fig:Config_Space}
	\end{figure}

Several alternatives and generalizations have already been proposed to address modifications of the zero-range contact interaction~\cite{PhysRevA.90.023620, PhysRevLett.82.2536, CHEON1998111,bougasImpactDarkStates2024}.
In this work, we consider a simple generalization of the standard contact interaction for two particles confined in an infinite box trap of length~$L$,
with coordinates $x_{1,2} \in [0,L]$. The decentered interaction potential is given by
    \begin{equation}
        V(g,c) = g \delta(x_2 - x_1 + c) + g \delta(x_2 - x_1 - c),
        \label{Eq:off_centered_Interaction}
    \end{equation}
where $g$ is the interaction strength and $c$ is the displacement 
parameter, as illustrated in Fig.~\ref{Fig:Config_Space}.

The interaction acts only when the particles are separated by a fixed distance $c$, which divides the space into three regions as shown in Fig.~\ref{Fig:Config_Space}.
Throughout this work, regions~I and~III are referred to as the region \textit{outside}, and region~II as the region \textit{inside}.
In the limit $c \to 0$, the model reduces to the conventional zero-range contact interaction.
The exchange symmetry $x_1 \leftrightarrow x_2$ ensures that the eigenstates can be classified as bosonic (symmetric) or fermionic (antisymmetric) under particle exchange.
However, $V$ acts away from $x_1 = x_2$, so antisymmetric wave functions need not vanish at its support. Therefore, this two-body interaction model, unlike the zero-range contact interaction, acts within both the bosonic and fermionic sectors.
In free space, $V(g,c)$ reduces, in the relative coordinate $r = x_2 - x_1$,
to the symmetric double Dirac delta potential $g[\delta(r-c)+\delta(r+c)]$, which is an exactly solvable scattering problem \cite{lapidus1982resonance,ahmed2016revisiting}. 

Previous work on an harmonically trapped system of two particles interacting via Eq.~(\ref{Eq:off_centered_Interaction}) has shown that the 
stationary properties of the system are strongly shaped by a class of noninteracting eigenstates that simultaneously solve the interacting problem \cite{bougasImpactDarkStates2024}, referred to as \emph{dark states},
in analogy to states in quantum optics that are insensitive to optical driving \cite{fleischhauer2005electromagnetically}.
Specifically, in the strong interaction limit $g \to \infty$, dark states mark points of triple degeneracy and set the boundaries between singly and doubly degenerate levels.
In addition, the structure of the low-energy spectrum is designed by the competition 
between $c$ and the harmonic oscillator length $a_{ho} = \sqrt{\hbar / m\omega}$. 
For $c <a_{ho}$, the system exhibits strong two-body exclusion, similar to the contact interaction model.
When $c >a_{ho}$, the tails of the wavefunctions are suppressed, which results in a particle bunching effect.
In the intermediate regime $c \sim a_{ho}$, exclusion and bunching compete, and small changes in $c$ can lead to dramatic changes in the wavefunction variance.
Dark states determine the competition between bunching and exclusion in the latter regime \cite{bougasImpactDarkStates2024}.

In contrast to a harmonic trap, where the system is integrable by the separability of the relative and center-of-mass coordinates, and exactly solvable for generic parameters, the box potential breaks separability and our system is not integrable for general parameters.
Nevertheless, dark states appear again, and remarkably, the subspace of the Hamiltonian they span remains exactly solvable for arbitrary parameters, leading to \emph{partial solvability}.
We remark that partial solvability has been found before in systems with two-body contact interactions \cite{10.21468/SciPostPhysCore.8.4.083, braakIntegrabilityWeakDiffraction2014, Fogarty2021probingedgebetween}.

The Hamiltonian for a two-body system with decentered contact interaction is given by
$H(g, c) = H_0 + V(g, c)$, where $V(g, c)$ is defined above and
  \begin{equation}
      H_0 = -\frac{\hbar^2}{2m}\left(\partial_{x_1}^2 + \partial_{x_2}^2\right).
      \label{Eq:Hamiltonian_Free}
  \end{equation}
Since the particles are trapped by hard walls, the wavefunction must satisfy Dirichlet boundary conditions,
  \begin{align*}
    \psi(x_1 = 0, x_2) = \psi(x_1, x_2 = L) = 0 
    \quad \text{if } x_2 \ge x_1,\\
    \psi(x_1 = L, x_2) = \psi(x_1, x_2 = 0) = 0 
    \quad \text{if } x_1 > x_2,
  \end{align*}
for all $x_1, x_2 \in [0, L]$. For general parameters, the system is solved via exact diagonalization,
with the Hamiltonian organized into symmetry-adapted blocks, see Sec.~\ref{sec:exact_diag}. Beyond the general case, four special
cases admit exact solutions: (i) the non-interacting limit where $g \to 0$; (ii) the
strong interaction limit $g \to \infty$ for arbitrary $c$, which is exactly solvable in the regions
outside; (iii) the centered limit $c \to 0$, which is tractable via the
Bethe ansatz; and (iv) the most physically rich case, in which $c$ takes certain rational
values, leading to dark states.

Notably, the dark states identified here bear a close resemblance to quantum many-body scars
(QMBS): non-thermal eigenstates embedded in a non-integrable system that span a decoupled
subspace of the Hilbert space not associated with any global symmetry of the Hamiltonian
\cite{serbynQuantumManybodyScars2021}. The dark states satisfy this definition precisely:
they are exact eigenstates of the full interacting Hamiltonian that retain the form of
non-interacting states and span a subspace that decouples from the rest of the spectrum
without being protected by any global symmetry.

The remainder of this article is organized as follows. 
In Section~\ref{sec:exact_diag}, we discuss the symmetries of the system and introduce the method of solution for general parameters. 
In Section~\ref{sec:special_cases}, we focus on three special cases. 
First, we consider the non-interacting limit of the problem and discuss the additional symmetries that emerge in this case.
Second, we focus on the zero-range (centered) contact interaction, 
where the Bethe ansatz provides exact analytical results.
Third, we extend the analysis to the strong interaction limit for arbitrary displacement $c$, and present the analytical solution for the region outside.
In Section~\ref{sec:DarkStates}, we identify and study dark states and discuss their scaling properties and the exactly solvable sector they define. 
We conclude in Section~\ref{sec:Conclusion} with a summary of our findings.

\section{Exact diagonalization for general parameters}
\label{sec:exact_diag}
We introduce scaled coordinates $\tilde{x}_i = x_i/L$ and measure energies in units of $E_0 = \hbar^2/(2mL^2)$.
Defining the dimensionless interaction strength and displacement,
    \begin{equation}
        \tilde{g} = \frac{2mgL}{\hbar^2}, \qquad \tilde{c} = \frac{c}{L},
        \label{eq:dimless_params}
    \end{equation}
the dimensionless Hamiltonian $\mathcal{H}(g,c) = H(g,c)/E_0$ takes the form
    \begin{equation}
        \mathcal{H} = 
        - \left(\partial_{\tilde{x}_1}^2 + \partial_{\tilde{x}_2}^2 \right)
        + \tilde{g}\left[\delta(\tilde{x}_2 - \tilde{x}_1 + \tilde{c}) 
        + \delta(\tilde{x}_2 - \tilde{x}_1 - \tilde{c})\right],
        \label{eq:hamiltonian_dimless}
    \end{equation}
with $\tilde{c}\in[0,1]$. 
For $\tilde{c}=1$, the $\delta$-barriers are placed at the boundaries of the box,
where all wavefunctions satisfy the Dirichlet boundary conditions, thus recovering the non-interacting limit.
Also, $\tilde{c}=0$ recovers the well-studied zero-range contact interaction.
Throughout this work, we drop tildes and work entirely in dimensionless units; all reported energies $\varepsilon_i = E_i/E_0$ are dimensionless.

We seek solutions to the dimensionless Schr\"odinger equation 
$\mathcal{H}\Psi_i(x_1,x_2) = \varepsilon_i\,\Psi_i(x_1,x_2)$, 
expanding the eigenfunctions as
    \begin{equation}
        \Psi_i(x_1,x_2) = \sum_{n,m} c_i^{nm}\,\psi_{nm}(x_1,x_2),
        \label{eq:expansion}
    \end{equation}
where the expansion coefficients $c_i^{nm}$ are determined by diagonalization.
The natural single-particle basis on $[0,1]$ is provided by the eigenfunctions of the infinite square well \cite{shankar2012principles},
    \begin{equation}
        \phi_n(x) = \sqrt{2}\sin(n\pi x), \quad n = 1, 2, 3, \ldots,
        \label{eq:spbasis}
    \end{equation}
with single-particle energies $\varepsilon_n^{(1)} = n^2\pi^2$.
An unsymmetrized complete set for two particles is given by the two-particle product states $\psi^{us}_{nm}(x_1,x_2) = \phi_n(x_1)\phi_m(x_2)$ with energies $\varepsilon_{nm}^{(2)} = (n^2 + m^2)\pi^2$.
The symmetrized basis $\psi_{nm}(x_1,x_2)$ is classified in Table~\ref{tab:symmetries}.

The Hamiltonian $\mathcal{H}(g,c)$ commutes with two symmetries for
arbitrary $g$ and $c$: particle permutation $\hat{\Sigma}$ and spatial inversion
$\hat{\Pi}$, acting as
    \begin{equation*}
        \hat{\Sigma} : (x_1, x_2) \to (x_2, x_1), \quad
        \hat{\Pi}    : (x_1, x_2) \to (1 - x_1,\, 1 - x_2).
    \end{equation*}
Together, $\hat{\Sigma}$ and $\hat{\Pi}$ generate the group $\mathcal{G} = \{\hat{e}, \hat{\Sigma}, \hat{\Pi}, \hat{\Pi}\hat{\Sigma}\}$,
isomorphic to the dihedral group $D_2$ (the symmetry group of a rectangle),
which is Abelian and has four one-dimensional irreducible representations.
In addition, the time-reversal operator $\hat{T}: t \mapsto -t$,
which for spin-0 bosons and spin-polarized fermions, where spin degrees of freedom are absent or frozen and do not enter the spatial Hamiltonian,
acts as $\hat{T}\psi(x_1,x_2) = \psi^*(x_1,x_2)$, also commutes with $\mathcal{H}(g,c)$.
Thus, the kinematic symmetry group  extends to $\mathcal{G} \times \{\hat{e}, \hat{T}\} \cong D_2 \times \mathcal{Z}_2$.
Time-reversal invariance implies that all eigenfunctions can be chosen real~\cite{sakurai2020modern}.
In the non-interacting limit, the symmetry group of the configuration space enlarges to $D_4$ (the symmetry group of a square),
which is non-Abelian; turning on the interaction reduces it back to $D_2$~\cite{harshmanOneDimensionalTrapsTwoBody2016}.

Since $\hat{\Sigma}^2 = \hat{\Pi}^2 = \hat{e}$, the eigenvalues of both operators are $\sigma, \pi = \pm 1$.
Consequently, the Hilbert space can be decomposed into four symmetry sectors labeled by the pair $(\sigma,\pi)$,
    \begin{equation}
        \mathcal{H} = \mathcal{H}_{+1,+1} \oplus \mathcal{H}_{+1,-1} \oplus \mathcal{H}_{-1,+1} \oplus \mathcal{H}_{-1,-1},
    \end{equation}
where each subspace $\mathcal{H}_{\sigma,\pi}$ contains states that are simultaneous eigenfunctions of $\hat{\Sigma}$ and $\hat{\Pi}$ with eigenvalues $\sigma$ and $\pi$, respectively.
For the basis states $\psi_{nm}(x_1,x_2)$ we have
    \begin{align}
        \hat{\Sigma} \psi_{nm}(x_1,x_2) &= \sigma \psi_{nm}(x_1,x_2), \\
        \hat{\Pi} \psi_{nm}(x_1,x_2)    &= (-1)^{n+m}\psi_{nm}(x_1,x_2).
    \end{align}
$\sigma = +1$ $(-1)$ corresponds to the bosonic (fermionic) sector.  
The parity $\pi = +1$ $(-1)$ indicates even (odd) total parity, determined by $n+m$.
The corresponding symmetry-adapted basis functions are listed in Table~\ref{tab:symmetries}.
    \begin{table}[t]
        \centering
        \caption{Symmetry-adapted two-particle basis functions $\psi_{nm}$,
                with $n \geq m$ for the bosonic sector ($\sigma = +1$) and $n > m$
                for the fermionic sector ($\sigma = -1$).}
        \label{tab:symmetries}
            \begin{ruledtabular}
                \begin{tabular}{ccl}
                    $\sigma$ & $\pi$ & \multicolumn{1}{c}{Basis function $\psi_{nm}$} \\
                    \hline
                    $+1$ & $+1$ &
                    $\dfrac{\phi_n(x_1)\phi_m(x_2)+\phi_m(x_1)\phi_n(x_2)}{\sqrt{2(1+\delta_{nm})}}$,\quad $n+m$ even \\[12pt]
                    $+1$ & $-1$ &
                    $\dfrac{\phi_n(x_1)\phi_m(x_2)+\phi_m(x_1)\phi_n(x_2)}{\sqrt{2(1+\delta_{nm})}}$,\quad $n+m$ odd  \\[12pt]
                    $-1$ & $+1$ &
                    $\dfrac{\phi_n(x_1)\phi_m(x_2)-\phi_m(x_1)\phi_n(x_2)}{\sqrt{2}}$,\quad $n+m$ even \\[12pt]
                    $-1$ & $-1$ &
                    $\dfrac{\phi_n(x_1)\phi_m(x_2)-\phi_m(x_1)\phi_n(x_2)}{\sqrt{2}}$,\quad $n+m$ odd  \\
                \end{tabular}
            \end{ruledtabular}
    \end{table}
    
Within each symmetry sector, the Hamiltonian matrix elements decompose as 
$\mathcal{H}_{nm,n'm'} = \mathcal{T}_{nm,n'm'} + \mathcal{V}_{nm,n'm'}$. 
The kinetic contribution is diagonal,
    \begin{equation}
        \mathcal{T}_{nm,n'm'} = \pi^2(n^2+m^2)\,\delta_{nn'}\delta_{mm'},
        \label{eq:kinetic_elements}
    \end{equation}
and the interaction matrix elements are
    \begin{multline}
        \mathcal{V}_{nm,n'm'} = g\int_0^1\!\int_0^1 
        \psi_{nm}^{*}
        V(g,c) 
        \psi_{n'm'}^{} dx_1 dx_2\\
        \quad = g \int_c^{1} 
        \psi_{nm}^{*}(x_1, x_1-c) 
        \psi_{n'm'}^{}(x_1, x_1-c) dx_1 \\
        \quad + g \int_0^{1-c} 
        \psi_{nm}^{*}(x_1, x_1+c) 
        \psi_{n'm'}^{}(x_1, x_1+c) dx_1.
        \label{eq:V_elements}
    \end{multline}
Details on the evaluation of Eq.~(\ref{eq:V_elements}) are given in Appendix~\ref{app:PotentialMatrix}.  

The matrix elements vanish when $n + m + n' + m' = 2k + 1$, with $k$ being an integer. This selection rule is a direct consequence of the total parity invariance of the potential:
the transition from even (odd) parity to odd (even) parity is forbidden.
Additionally the interaction potential commutes with the exchange operator $\hat{\Sigma}$: for any fermionic state $|\psi^F_{nm}\rangle$ and bosonic state $|\psi^B_{nm}\rangle$, we have  $\langle \psi^F_{nm} | \hat{V} | \psi^B_{nm} \rangle = 0$.
The exchange symmetry invariance of the potential also implies $V_{nm,n'm'} = V_{mn,m'n'}$.

To solve the two-body interacting problem for general parameters, we employ exact diagonalization,
independently within each symmetry sector.
The basis functions $\psi_{nm}$ are indexed by single-particle wavenumbers
$n,m \leq n_{\text{max}}$, which gives a total basis size
    \begin{equation}
        N_{\text{basis}} = \frac{n_{\text{max}}(n_{\text{max}} \pm 1)}{2},
    \end{equation}
for bosons $(+)$ and fermions $(-)$ respectively, in the absence of additional symmetries.
We impose the energy cut-off $n^2+m^2 \leq n_{\text{max}}^2$,
to exclude high-energy states whose contribution to the low-lying spectrum is negligible.
Note that the choice of $n_{\text{max}}$ balances accuracy and computational cost; in all
computations we use $n_{\text{max}} = 320$.

To obtain the lowest-lying eigenvalues and eigenstates, we employ the implicitly restarted Lanczos algorithm \cite{lanczos1950iteration} as implemented in ARPACK \cite{lehoucq1998arpack} and accessed via SciPy's \texttt{eigsh} \cite{virtanen2020scipy}. 
The Lanczos method is a special case of the Arnoldi algorithm \cite{arnoldi1951principle}.
It exploits the Hermiticity of the Hamiltonian to project it onto a Krylov subspace,
yielding a real symmetric tridiagonal matrix whose eigenvalues approximate the extremal eigenvalues (in our case the lowest-lying eigenvalues) of the original Hamiltonian \cite{Weiße2008}.
This iterative approach allows one to compute a few extremal eigenvalues without the need for performing a full diagonalization of the Hamiltonian matrix.
The ED results are validated in the next section by comparison with analytic results obtained for special limits. 

    \begin{figure*}[htbp]
        \centering
        \includegraphics[width= 0.8 \linewidth]{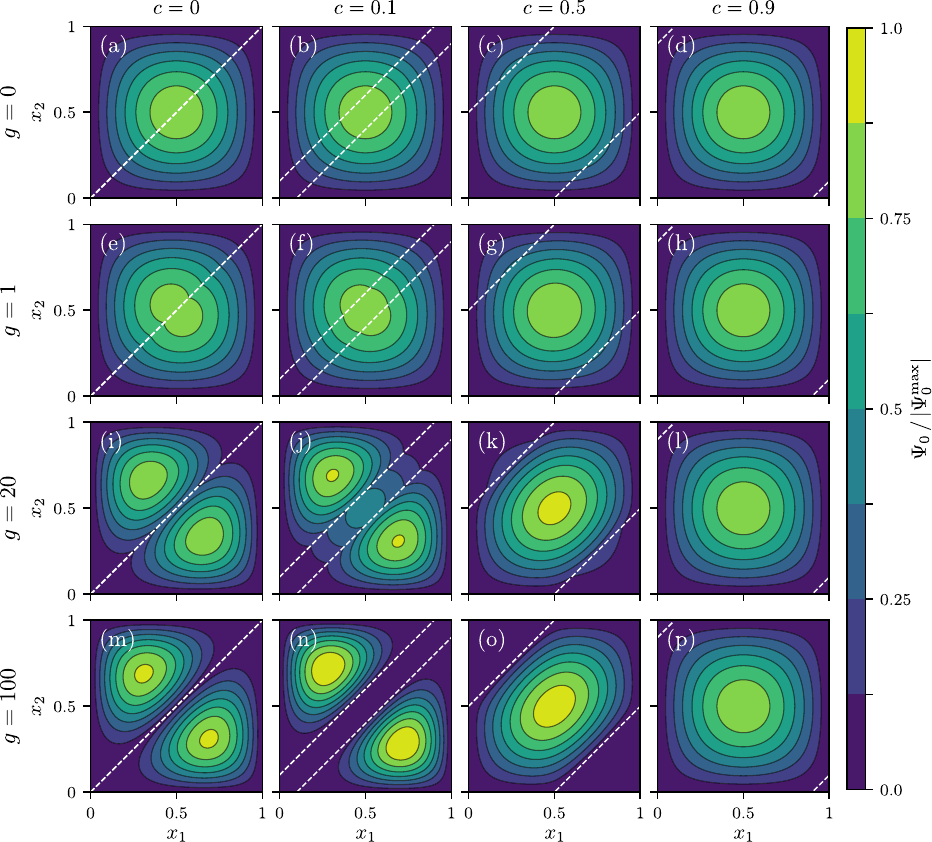}
        \caption{The ground-state wave function $\psi(x_1,x_2)$ in
                the bosonic even-parity sector $(\sigma,\pi)=(+1,+1)$, for interaction
                strengths $g = 0, 1, 20, 100$ (rows, top to bottom) and displacement parameters
                $c = 0, 0.1, 0.5, 0.9$ (columns, left to right).
                The dashed lines indicate the support of the interaction potential.
                At $c = 0$, the model reduces to the standard contact interaction.}

        \label{Fig:Ground_Multi_Sym=SE}
    \end{figure*}

Figure~\ref{Fig:Ground_Multi_Sym=SE} shows the ground state wavefunction in the bosonic even-parity sector for a grid of interaction strengths $g$ and displacement parameters $c$; where results are obtained entirely using the discussed ED approach.
At $c=0$, the model reduces to the standard contact interaction.
For $c=0.1$, increasing $g$ suppresses the wavefunction in the region inside, and the wavefunction is mainly supported in the region outside.
In contrast, for $c=0.5$, the wavefunction tends to localize in the region inside as $g$ increases.
For $c=0.9$, the interaction lines lie very close to the box boundaries,
and the particles barely feel the interaction for any value of $g$;
the wavefunction remains close to the non-interacting ground state of two particles in a hard-wall box for all values of $g$ shown.
\section{Three special cases}
\label{sec:special_cases}

\subsection{The non-interacting limit \texorpdfstring{$g \to 0$}{}}
\label{subsec:Non-Interacting_Limit}
In the non-interacting limit, the Hamiltonian reduces to $\mathcal{H}_0 = -\partial_{x_1}^2 - \partial_{x_2}^2$,
and the two-particle Schrödinger equation separates into two independent single-particle problems, each of the form 
$-\partial_{x_i}^2 \phi_n(x_i) = \varepsilon_n \phi_n(x_i)$, with $\varepsilon_n = n^2\pi^2$.
The symmetry-adapted eigenstates are those already constructed in Table~\ref{tab:symmetries}.
The non-interacting system is thus integrable by separability: each quantum number $n, m$ is individually conserved,
and the full spectrum can be obtained analytically from the single-particle spectrum without the need for diagonalization. 

Let $(n, m)$ be a primitive solution with $\gcd(n,m)=1$.
New eigenstates can be generated from this elementary solution via \emph{integer scaling} by a single integer $k$.
More specifically,
the set
    \begin{equation}
        \bigl\{(kn, km) \big| k \in \mathbb{Z} \bigr\},
    \end{equation}
defines a tower of states, i.e., a family of eigenstates with a common ray in the quantum-number plane at polar angle $\theta_{nm}=\arctan(m/n)$,
in which the energy of the $k$-th member is $\varepsilon_{kn, km} = k^2 \varepsilon_{nm}$.
Within a tower, the symmetry under particle permutation $\sigma$ is always preserved for all $k$.
The total parity under spatial inversion of the $k$-th member is $\pi_k = (-1)^{k(n+m)}.$
If the primitive state is even under spatial inversion ($n+m$ even), then $\pi_k=+1$ for all $k$.
If the primitive state is odd under spatial inversion ($n+m$ odd), then $\pi_k=(-1)^k$: odd-$k$ members remain odd under spatial inversion ($\pi=-1$),
while even-$k$ members become even ($\pi=+1$).

More generally, however, there exist scale-rotation transformations that connect solutions across different towers,
which also provides a procedure for analyzing accidental degeneracies rooted in number theory rather than kinematic symmetries.
The corresponding `\emph{step-up}' operators $J_{a,b}$, with $a,b \in \mathbb{Z}$ act as geometric transformations in the quantum-number plane $J_{a,b}:\ (n,m) \mapsto (n',m') = (an - bm, am + bn)$, which correspond to scaling by $\sqrt{a^2 + b^2}$  and a rotation by $\theta_{ab} = \tan^{-1}(b/a)$. Under this transformation, the energy rescales as $\varepsilon_{n',m'} = (a^2 + b^2) \varepsilon_{n,m}$. It can be shown that the set of operators $\{J_{a,b}\}$ forms a commuting ring, rather than a group~\cite{GBShaw_1974}.

The degeneracy structure in the spectrum of a given system can be mostly explained by the 
dimension of the irreducible representations (irreps) of its kinematic symmetry group.
For two non-interacting particles, in general, the minimal kinematic symmetry is  $S_2 \ltimes \mathcal{K}_1^{\times 2} \subseteq \mathcal{K}_2^{(0)}$,
where $S_2$ is the permutation group of two particles generated by $\hat{\Sigma}$, 
$\mathcal{K}_1$ is the kinematic symmetry group of a single particle,
$\ltimes$ denotes the semidirect product with $S_2$ acting as a non-trivial automorphism on $\mathcal{K}_1^{\times 2}$,
and $\mathcal{K}_2^{(0)}$ is the full kinematic symmetry group of the non-interacting two-particle system \cite{harshmanOneDimensionalTrapsTwoBody2016}.
Specifically, for a symmetric trap which respects parity, $\mathcal{K}_1 \sim \mathcal{Z}_2 \times \mathcal{T}_t$,
where $\mathcal{Z}_2$ is the cyclic group generated by the spatial inversion in one-particle configuration space, and $\mathcal{T}_t$ is the group of unitary time 
translation operators generated by $\hat{U}_i(t) = \exp(-i \mathcal{H}_i t)$ where $\mathcal{H}_i$ is the Hamiltonian of each single particle.
Since each particle carries its own copy of $\mathcal{Z}_2 \times \mathcal{T}_t$, the full minimal kinematic 
symmetry of the non-interacting two-particle system is $S_2 \wr (\mathcal{Z}_2 \times \mathcal{T}_t)$, where the wreath product structure captures the 
independent action of $\mathcal{Z}_2 \times \mathcal{T}_t$ on each particle together with 
their permutation \cite{harshmanIdenticalWellsSymmetry2017a, PhysRevA.95.053616}.

The group $\mathcal{Z}_2 \times \mathcal{T}_t$ is abelian,
and its irreps are one-dimensional labeled by pairs $(\pi, \varepsilon_n)$,
where $\pi = \pm 1$ denotes parity and $\varepsilon_n$ the single-particle energy level.
The irreps of $S_2 \wr (\mathcal{Z}_2 \times \mathcal{T}_t)$ then fall into two classes:
when the two particles carry distinct quantum numbers $(\pi,\varepsilon_n)\neq(\pi',\varepsilon_{n'})$,
the corresponding states 
are mapped into each other 
by particle exchange but are not individually preserved,
so no smaller invariant subspace exists: they form a single irreducible two-dimensional representation.
When both particles share the same quantum numbers,
each two-particle state is individually invariant under $S_2$,
and the two irreps of $S_2$ yield two distinct one-dimensional irreps.
This symmetry group is thus sufficient to explain \textit{all} double degeneracies in the non-interacting spectrum,
with one exception: the \textit{Pythagorean degeneracies}, which do not originate from any symmetry in the Hamiltonian, and are therefore regarded as accidental degeneracies \cite{10.1119/1.18734, GBShaw_1974}. 

\subsection{Bethe-ansatz solution for \texorpdfstring{$c = 0$}{}}
\label{subsec:contact_limit}
In this section, we study the special case of zero-range contact interaction.
In a hard-wall box trap, this model is Bethe-ansatz integrable and has been solved exactly for both bosonic and fermionic sectors ~\cite{Batchelor_2005, Oelkers_2006}. 
In this case, the fermionic spectrum is insensitive to the interaction strength $g$, thus $\varepsilon_{nm}^{(f)} = \pi^2(n^2 + m^2)$.
In the bosonic sector, the eigenenergies are determined by two decoupled transcendental equations for the center-of-mass quasi-momentum $K = k_1+k_2$ and the relative quasi-momentum $\Delta = k_1-k_2$,
    \begin{subequations}
    \label{eq:bethe_main}
        \begin{align}
            K_{nm}      &= 2\arctan\!\left(\frac{g}{K_{nm}}\right) + \pi(n + m),
            \label{eq:bethe_K_main}\\[4pt]
            \Delta_{nm} &= 2\arctan\!\left(\frac{g}{\Delta_{nm}}\right) + \pi(n - m),
            \label{eq:bethe_Delta_main}
        \end{align}
    \end{subequations}
with $n \geq m \geq 1$ and eigenenergy $\varepsilon^{(b)}_{n m} = (\Delta_{nm}^2+K_{nm}^2)/2$.
Note that this decoupled form is unique to a two-body system.
The full derivation is given in Appendix~\ref{app:bethe}.
In the non-interacting limit ($g\to 0$),
one recovers $\varepsilon^{(b)}_{nm} = \pi^2(n^2 + m^2)$.  In the $g\to\infty$ limit, $\arctan(g/q)\to\pi/2$, 
thus Eqs.~(\ref{eq:bethe_main}) reduce to $K_{nm} = \pi(n+m+1)$ and $\Delta_{nm} = \pi(n-m+1)$. 
Hence, the resulting spectrum is $\varepsilon^{(b)}_{nm} = \pi^2\bigl((n+1)^2+m^2\bigr)$ which is identical to that of two free fermions.
In this limit, the Girardeau mapping~\cite{girardeau1960relationship},
$\psi^{g\to\infty}_{\mathrm{bos}} =\bigl|\psi^{g=0}_{\mathrm{fer}}\bigr|$, is recovered, and the bosonic system is called to be fermionized.
Additionally, every bosonic eigenstate becomes degenerate with its fermionic counterpart of opposite parity, see Sec.~\ref{subsec:inf_g}.

The exact solution follows from the Bethe-ansatz integrability of the 
zero-range contact interaction model. Although the total momentum is not conserved in 
the hard-wall box (since reflections from the walls send $k_j \to -k_j$) both elastic two-body scattering and reflection preserve the integrals of motion $I_1 = k_1^2 + k_2^2$ and $I_2 = k_1^4 + k_2^4$.

     \begin{table}[t]
        \centering
        \caption{Comparison of Bethe ansatz and ED ground-state energies for 
        $c = 0$ at $g = 1, 20, 100$. The exchange symmetry $\sigma = \pm 1$ 
        and spatial parity $\pi = \pm 1$ label the four sectors. The relative 
        error is defined as $|\varepsilon_\mathrm{ED} - \varepsilon_\mathrm{Bethe}|/\varepsilon_\mathrm{Bethe}$.}
        \label{tab:bethe_ed}
            \begin{ruledtabular}
                \begin{tabular}{cccrrr}
                    $g$ & $\sigma$ & $\pi$ & 
                    \multicolumn{1}{c}{$\varepsilon_\mathrm{Bethe}$} & 
                    \multicolumn{1}{c}{$\varepsilon_\mathrm{ED}$} & 
                    \multicolumn{1}{c}{Rel.\ error} \\
                    \hline
                    \multirow{4}{*}{$1$}
                    & $+1$ & $+1$ & 22.53213 & 22.53329 & $5.15\times10^{-5}$ \\
                    & $+1$ & $-1$ & 53.13085 & 53.13245 & $3.01\times10^{-5}$ \\
                    & $-1$ & $+1$ & 98.69604 & 98.69604 & $ = 0$         \\
                    & $-1$ & $-1$ & 49.34802 & 49.34802 & $= 0$         \\[4pt]
                    \multirow{4}{*}{$20$}
                    & $+1$ & $+1$ & 41.16319 & 41.22145 & $1.42\times10^{-3}$ \\
                    & $+1$ & $-1$ & 82.75898 & 82.86726 & $1.31\times10^{-3}$ \\
                    & $-1$ & $+1$ & 98.69604 & 98.69604 & $ = 0$         \\
                    & $-1$ & $-1$ & 49.34802 & 49.34802 & $ = 0$         \\[4pt]
                    \multirow{4}{*}{$100$}
                    & $+1$ & $+1$ & 47.43657 & 47.51929 & $1.74\times10^{-3}$ \\
                    & $+1$ & $-1$ & 94.87942 & 95.04446 & $1.74\times10^{-3}$ \\
                    & $-1$ & $+1$ & 98.69604 & 98.69604 & $ = 0$         \\
                    & $-1$ & $-1$ & 49.34802 & 49.34802 & $ = 0$         \\
                \end{tabular}
            \end{ruledtabular}
    \end{table}

We compare the exact solution from Eqs.~\eqref{eq:bethe_main} with the results obtained by our ED approach for $c = 0$ to benchmark our ED approach.
Table~\ref{tab:bethe_ed} compares the ground state energies obtained via Bethe ansatz and ED for $g \in \{1, 20, 100\}$ across all symmetry sectors.
The fermionic sectors are insensitive to the interaction strength; thus, they are reproduced exactly by ED at all $g$. The bosonic sectors show a small but growing relative error with increasing $g$, reaching at 
most $\sim 10^{-3}$ at $g = 100$. The excellent agreement across all sectors and interaction strengths 
validates the ED approach as a reliable method to obtain solutions where analytical methods do not exist. 

\subsection{Infinite \texorpdfstring{$g$}{g}, arbitrary \texorpdfstring{$c$}{c}}
\label{subsec:inf_g}
In the limit $g \to \infty$, the delta interactions become impenetrable barriers, thus,
they impose the constraint that the wavefunction must vanish at $|x_2 - x_1| = c$.
The configuration space is then divided into three disconnected regions: Region~I ($x_2 - x_1 > c$),
Region~II ($-c \leq x_2 - x_1 \leq c$), and Region~III ($x_2 - x_1 < -c$).
Within each region, the two-particle Schr\"{o}dinger equation reduces
to a free 2D Helmholtz equation with Dirichlet boundary conditions on the walls of the polygonal domain,
equivalent to the problem of quantum billiards in polygons~\cite{liboff1994polygon, richens1981pseudo, gutkin1986,zyczkowskiClassicalQuantumBilliards, lozejIntermediateSpectralStatistics2024}.
In this limit, the time-translation symmetry is parametrized by $t_R$ and represented by
    \begin{equation}
        \hat{U}_R(t_R) = \exp(-i \mathcal{H}_R t_R), \qquad R \in {\mathrm{I}, \mathrm{II}, \mathrm{III}}.
        \label{eq:unitary_regions}
    \end{equation}
within each region.
Equivalently, the phase difference between disjoint regions in the $g\to\infty$ limit is not an observable quantity~\cite{PhysRevA.95.053616}.

Regions~I and III each correspond to a right isosceles triangular
domain of leg length $a = 1 - c$ in the two-particle configuration space;
this triangular quantum billiard is integrable and exactly solvable~\cite{zyczkowskiClassicalQuantumBilliards, pandaClassicalPeriodicOrbits2019b}.
The corresponding eigenfunctions are
    \begin{multline}
        \Psi_{nm}(x_1, x_2) = \frac{2}{a} [\sin(k_n x_1)\sin(k_m x_2) \\
        \quad - \sin(k_m x_1)\sin(k_n x_2)],
        \label{eq:triangle_eigenfunctions}
    \end{multline}
where $k_n = \pi n/a$ and $k_m = \pi m/a$ are the quantized quasimomenta, with
energies $\varepsilon_{nm} = k_n^2 + k_m^2 = \pi^2(n^2 + m^2)/a^2$, $n > m \geq 1$.

    \begin{figure}[t]
        \centering
        \includegraphics[width= \linewidth]{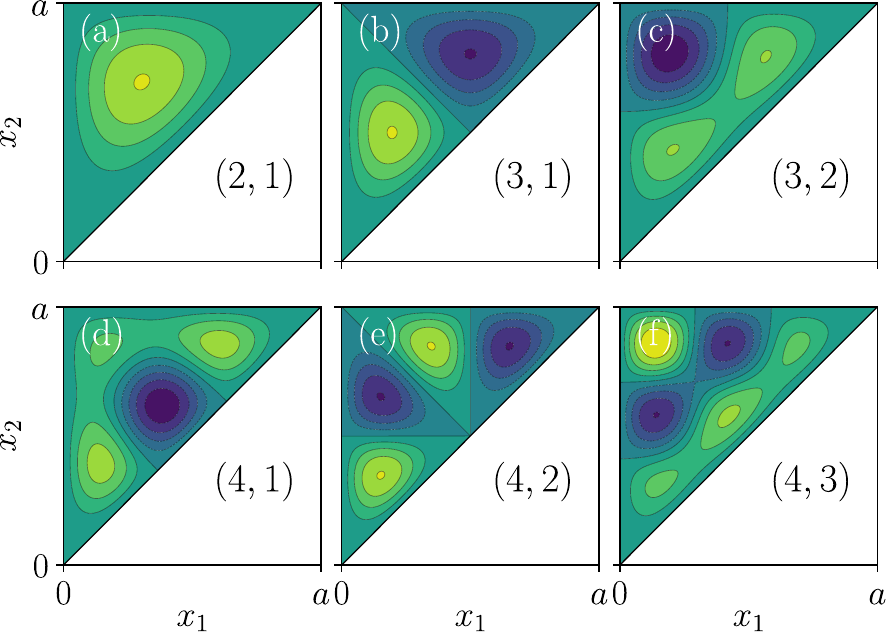}
        \caption{The six lowest-lying eigenfunctions $\Psi_{nm}(x_1, x_2)$ on the right isosceles 
                triangular domain of leg length $a$. Quantum numbers $(n,m)$ are given in the figure.}
        \label{Fig:Triangles}
    \end{figure}

The lowest eigenfunctions in Eq.~(\ref{eq:triangle_eigenfunctions}) are depicted in Fig.~\ref{Fig:Triangles}. 
The solution in region~I and III can be constructed via the Girardeau mapping~\cite{girardeau1960relationship}:
in the limit $g\to\infty$ the bosonic wavefunction satisfies $\psi^{g\to\infty}_\mathrm{bos} = |\psi^{g=0}_\mathrm{fer}|$,
mapping the problem onto that of two non-interacting fermions confined to a box of length $a$,
restricted to the domain $0 \leq x_1, x_2 \leq a$. The fermionic wavefunction for two non-interacting particles in a box of length $a$ is then given by
antisymmetric combination of the separable single-particle solutions, which yields Eq.~(\ref{eq:triangle_eigenfunctions}).
The quasimomenta are quantized via the Bethe ansatz equations for free fermions with hard-wall boundaries~\cite{Oelkers_2006}
    \begin{equation}
        e^{2ik_j a} = 1, \qquad j = 1, 2.
        \label{eq:bethe_hard_wall}
    \end{equation}
    \begin{figure*}[t]
        \centering
        \includegraphics[width= 0.95 \linewidth]{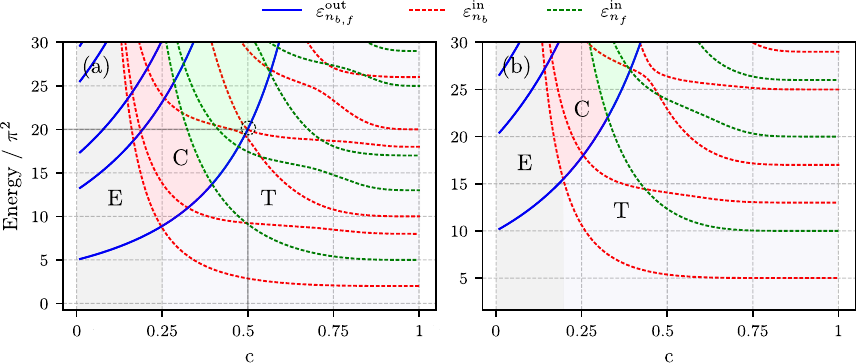}
        \caption{Energy levels at $g \to \infty$. (a) Bosonic even-parity and fermionic odd-parity sectors. (b) Bosonic odd-parity and fermionic even-parity sectors.
        $\varepsilon^{\text{out}}_{n_{b,f}}$ denotes the doubly degenerate fermionic and bosonic energy levels of the region outside.
        $\varepsilon^{\text{in}}_{n_b}$ and $\varepsilon^{\text{in}}_{n_f}$ denote the non-degenerate levels of the region inside for bosonic and fermionic states, respectively.
        The exact crossings between the lowest-energy levels $\varepsilon^{\text{out}}_{0_{b,f}}$ and
        $\varepsilon^{\text{in}}_{0_{b,f}}$ define four regimes:
        (E) exclusion, with only $\varepsilon^{\mathrm{out}}_{n_{b,f}}$; (T) truncation, with only $\varepsilon^{\mathrm{in}}_{n_{b,f}}$;
        and two crossover regions (C) where they coexist; red: $\varepsilon^{\mathrm{out}}_{n_{b}}$ with $\varepsilon^{\mathrm{in}}_{n_b}$; green: $\varepsilon^{\mathrm{out}}_{n_{b,f}}$ with both $\varepsilon^{\mathrm{in}}_{n_f}$, $\varepsilon^{\mathrm{in}}_{n_b}$.
        The triple degeneracy indicated by the dashed circle in panel (a) corresponds to the lowest-lying dark state (see Sec.~\ref{sec:DarkStates}).}  
        \label{Fig:Combined_EvsC_g_infty}
    \end{figure*}
For each particle ordering in regions~I and III, we define the \emph{snippet} basis wavefunction \cite{harshmanOneDimensionalTrapsTwoBody2016, PhysRevLett.100.160405, PhysRevA.84.023626} in our original coordinates,
i.e.,\ $0 \leq x_1, x_2 \leq 1$ by
    \begin{subequations}
    \label{eq:snippets}
        \begin{align}
            \psi^{\mathrm{I}}_{nm}(x_1,x_2) &=
                \begin{cases}
                    \Psi_{nm}(x_1, x_2 - c) & x_1 < x_2, \\
                    0                          & x_1 > x_2,
                \end{cases}
                \label{eq:domainI} \\
            \psi^{\mathrm{III}}_{nm}(x_1,x_2) &=
                \begin{cases}
                    0                            & x_1 < x_2, \\
                    -\Psi_{nm}(x_2, x_1 - c)  & x_1 > x_2,
                \end{cases}
                \label{eq:domainIII}
        \end{align}
    \end{subequations}
which satisfy $\hat{\Sigma} \psi^{\mathrm{I}}_{nm} = \psi^{\mathrm{III}}_{nm}$ under particle exchange.
From these, we construct the complete symmetric and antisymmetric basis
states spanning the full \textit{outside} domain,
    \begin{align}
        \Phi^{(+)}_{nm}(x_1,x_2) &= \frac{1}{\sqrt{2}}
            \bigl[\psi^{\mathrm{I}}_{nm} + \psi^{\mathrm{III}}_{nm}\bigr],
        \label{eq:sym_state} \\
        \Phi^{(-)}_{nm}(x_1,x_2) &= \frac{1}{\sqrt{2}}
            \bigl[\psi^{\mathrm{I}}_{nm} - \psi^{\mathrm{III}}_{nm}\bigr].
        \label{eq:antisym_state}
    \end{align}
Under the spatial inversion operator $\hat{\Pi}$ these states transform as
    \begin{align}
        \hat{\Pi}\,\Phi^{(+)}_{nm} &= -(-1)^{n+m}\Phi^{(+)}_{nm},
        \label{eq:parity_sym} \\
        \hat{\Pi}\,\Phi^{(-)}_{nm} &= \phantom{-}(-1)^{n+m}\Phi^{(-)}_{nm},
        \label{eq:parity_antisym}
    \end{align}
so that symmetric and antisymmetric states always carry opposite parity
eigenvalues. Since both states share the same energy eigenvalue,
the spectrum in the region outside exhibits a two-fold degeneracy in the $g \to \infty$ limit: 
every state in the fermionic (bosonic) symmetry sector of the Hamiltonian is degenerate with a corresponding state in the bosonic (fermionic) sector of opposite parity.

The system in the region outside, as in the non-interacting case,
is exactly solvable by separation of variables,
and the kinematic symmetry group is again $S_2 \wr (\mathcal{Z}_2 \times \mathcal{T}_t)$.
Regions I $\&$ III independently admit a $\mathcal{Z}_2 \times \mathcal{T}_t$ symmetry:
the $\mathcal{T}_t$ within each is generated by the corresponding time-evolution operator, see Eq.~(\ref{eq:unitary_regions});
and the $\mathcal{Z}_2$ is generated by
    \begin{equation}
        \hat{\rho}_{\alpha} =
        \bigl(\hat{\Pi}\,\hat{\Sigma}\bigr)\big|_{\alpha},
        \qquad \alpha\in\{\mathrm{I},\mathrm{III}\}.
    \end{equation}
$\hat{\rho}_\alpha$ acts as $(x_1,x_2)\mapsto(1-x_2,\,1-x_1)$, \emph{locally} in the region $\alpha$ and act as an identity in the other region.
$\hat{\rho}_\alpha$ preserves the particle ordering within each domain and satisfies
$\hat{\rho}_{\alpha}^2 = 1$.
The particle-exchange operator $\hat{\Sigma}$ interchanges both generator pairs
simultaneously,
    \begin{equation}
        \hat{\Sigma} \hat{U}_{\mathrm{I}}(t) = \hat{U}_{\mathrm{III}}(t) \hat{\Sigma},
        \qquad
        \hat{\Sigma} \hat{\rho}_{\mathrm{I}} = \hat{\rho}_{\mathrm{III}} \hat{\Sigma},
        \label{eq:intertwine}
    \end{equation}
so it permutes the two $\mathcal{Z}_2\times\mathcal{T}_t$ factors rather than
commuting with each individually. The kinematic symmetry group of the outer
regions is therefore
    \begin{equation}
        S_2\wr\bigl(\mathcal{Z}_2\times\mathcal{T}_t\bigr)
        \cong
        \bigl(\mathcal{Z}_2\times\mathcal{T}_t\bigr)^{\times 2}\rtimes S_2.
    \end{equation}

Region~II has interior angles $\pi/2$ at $(0,0)$ and $(1,1)$, and $3\pi/4$ at
its four remaining vertices, and tessellates the plane under translation.
A polygonal billiard is classically integrable only if every interior
angle is $\pi/n$ ($n\in\mathbb{Z}^+$)~\cite{gutkin1986}.
Only four such billiards exist: the rectangle and the triangles with angles
$(\pi/4,\,\pi/4,\,\pi/2)$, $(\pi/3,\,\pi/3,\,\pi/3)$, and
$(\pi/6,\,\pi/3,\,\pi/2)$.
The classical motion for these can be mapped onto a closed surface of genus one:
their phase space has the topology of a torus.
These polygons are also quantum integrable~\cite{liboff1994polygon} and admit a
complete set of trigonometric eigenfunctions~\cite{mccartin2008polygonal}.
All four polygons furthermore tessellate the plane under reflection.
However, polygons whose angles are rational multiples of $\pi$ and that
tessellate the plane under translation are termed
pseudointegrable~\cite{richens1981pseudo}; solving their quantum billiards
generically requires numerical methods~\cite{zyczkowskiClassicalQuantumBilliards}.
Region~II therefore falls into the pseudointegrable class, and we solve it numerically
via exact diagonalization discussed in Sec.~\ref{sec:exact_diag}.

Figure~\ref{Fig:Combined_EvsC_g_infty} shows the energy spectrum in the $g \to \infty$
limit as a function of $c$.
Particles occupying bosonic and fermionic eigenstates with energies $\varepsilon^{\mathrm{out}}_{n_{b,f}}$ are
localized in the region outside, while those with energies $\varepsilon^{\mathrm{in}}_{n_{b,f}}$
are localized within the region inside.
These two classes of eigenstates belong to different Hamiltonians,
thus all crossings between $\varepsilon^{\mathrm{out}}$ and $\varepsilon^{\mathrm{in}}$ are exact.
Varying the displacement $c$ is equivalent to translating the
$\delta$-barriers toward the hard-wall boundaries of the box; as $c\to 1$, the
deltas approach the hard walls, and the two particles become effectively noninteracting.

As in the harmonic potential case~\cite{bougasImpactDarkStates2024}, the lowest
$\varepsilon^{\mathrm{out}}$ and $\varepsilon^{\mathrm{in}}$ levels define, as a
function of $c$, distinct regimes that neatly classify the density profiles of the
two-particle system.
We define $c^{\dagger}$ as the exact crossing between the lowest-energy levels
$\varepsilon^{\mathrm{out}}_{0_{b,f}}$ and $\varepsilon^{\mathrm{in}}_{0_{b}}$.
In the \emph{exclusion} regime~(E), $c \ll c^{\dagger}$: only
$\varepsilon^{\mathrm{out}}_{n_{b,f}}$ levels are populated, all eigenstates are
doubly degenerate, each pair consisting of a bosonic and a fermionic state from
opposite parity sectors, and both particles reside in the region outside.
In the \emph{truncation} regime~(T), $c \gg c^{\dagger}$: only
$\varepsilon^{\mathrm{in}}_{n_{b,f}}$ levels are present, eigenstates are
generically nondegenerate (except for accidental degeneracies), and the particles
are localized in the region inside.
Between these two regions, two \emph{crossover} regimes~(C)
exist in which $\varepsilon^{\mathrm{out}}$ and $\varepsilon^{\mathrm{in}}$ levels
coexist: in the red crossover, $\varepsilon^{\mathrm{out}}_{n_{b,f}}$ coexists
with only the bosonic inner levels $\varepsilon^{\mathrm{in}}_{n_b}$; in the green
crossover, $\varepsilon^{\mathrm{out}}_{n_{b,f}}$ coexists with both
$\varepsilon^{\mathrm{in}}_{n_b}$ and $\varepsilon^{\mathrm{in}}_{n_f}$.
In both crossover regimes, nondegenerate and doubly degenerate eigenstates are
simultaneously present.
Dark states occur at points of triple degeneracy and therefore reside in the
crossover regime, as in the harmonic potential case.
However, unlike the harmonic potential, not every triple degeneracy corresponds
to a dark state; only a specific subset does, as discussed in
Sec.~\ref{sec:DarkStates}.

To validate our ED approach, we also compare its predictions against the analytical
results for the region outside in the $g \to \infty$ limit.
We evaluate the mean relative error of the ground-state energy,
    \begin{equation}
        \bar{\delta} = \frac{1}{N}\sum_{i=1}^{N}
        \frac{|\varepsilon^{\mathrm{ED}}_0(c_i) - \varepsilon^{\mathrm{Bethe}}_0(c_i)|}
            {\varepsilon^{\mathrm{Bethe}}_0(c_i)},
    \end{equation}
averaged over $N$ values of $c$ in the domain shown in Fig.~\ref{Fig:Combined_EvsC_g_infty}, for each
symmetry sector.
The resulting errors are $0.00149$ (bosonic, even parity), $0.00128$ (bosonic, odd parity), $0.00121$ (fermionic, even parity), and
$0.00150$ (fermionic, odd parity).

\section{Dark States}
\label{sec:DarkStates}
\begin{figure*}[t]
        \centering
        \includegraphics[width= 0.8 \linewidth]{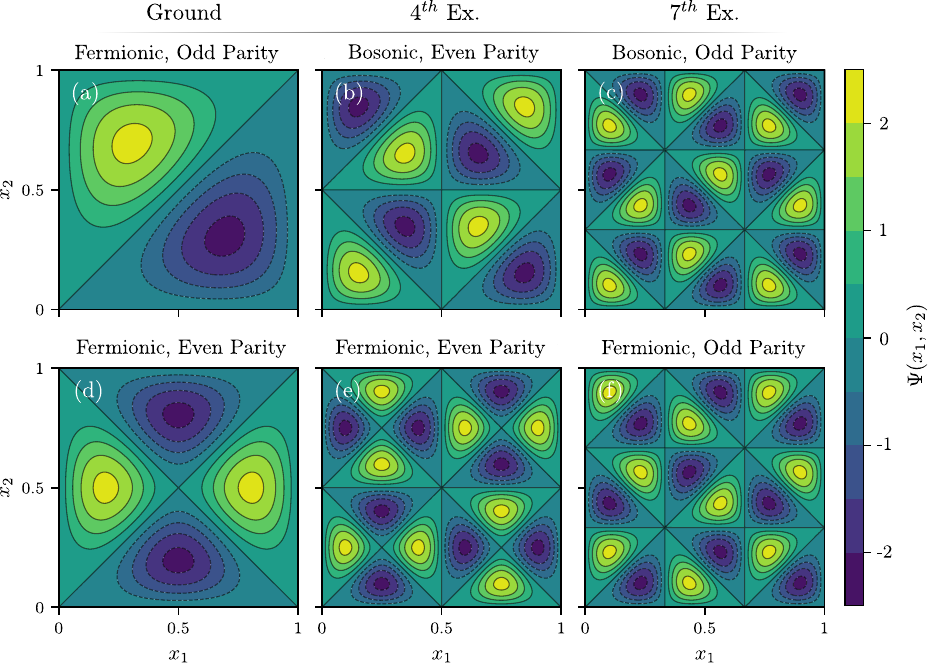}
        \caption{Wavefunctions of the lowest-lying $g=0$ eigenstates with
        diagonal nodal lines. The corresponding symmetry sector is shown in the figure.
        (a) and (d) carry diagonal nodal lines only at $c=0$, and are therefore
        dark at $c=0$.
        (b) and (e)~States that become dark at $c=1/2$.
        (c)~State that becomes dark at $c=1/3$.
        (f)~State that becomes dark at $c=2/3$.}
        \label{Fig:DarkStates}
    \end{figure*}

As discussed in Sec~\ref{subsec:Non-Interacting_Limit}, in the non-interacting limit $g \to 0$,
the two-particle eigenfunctions are symmetrized (fermionic or bosonic) combinations of single-particle box states.
A subset of these eigenfunctions possesses nodal lines parallel to the $\delta$-barriers $x_2 = x_1 \pm c$,
as illustrated in Fig.~\ref{Fig:DarkStates}.
This subset is of particular importance, as it gives rise to \emph{dark states} when the displacement $c$ takes specific values.
A state $\Psi$ is dark if $V \Psi = 0$,
which requires
    \begin{equation}
        \Psi(x_1, x_1 \pm c) = 0.
        \label{eq:dark_condition}
    \end{equation}
Dark states lie in the null space of $V(g,c)$ and are therefore insensitive to the interaction for arbitrary $g$. 

Fig.~\ref{Fig:DarkStates}, illustrates the lowest-lying wavefunctions that exhibit this diagonal nodal structure,
and therefore give rise to dark states for specific values of $c$.
Panels~(a) and~(d) give dark state to the interaction when $c = 0$: here both dark states 
belong to the fermionic sector, consistent with the well-known result that the conventional zero-range contact interaction ($c = 0$) leaves the fermionic sector completely unperturbed.
Panels~(b) and~(e) give rise to a dark state for $c = 1/2$, panel~(c) for 
$c = 1/3$, and panel~(f) for $c = 2/3$. These lowest-lying dark states for $c > 0$ and their 
quantum numbers are summarized in Table~\ref{tab:dark_states}.

    \begin{table}[htbp]
	    \caption{Lowest-lying dark states for $c > 0$ with energies $\varepsilon_{nm} = \pi^2(n^2 + m^2)/(1-c)^2$.
	    The corresponding nodal structures are illustrated in Fig.~\ref{Fig:DarkStates}.}
	    \label{tab:dark_states}
	        \begin{ruledtabular}
	            \begin{tabular}{ccccc}
	            $c$ & $(n, m)$ & $\varepsilon_{nm} / \pi^2$ & Symmetry sector & Panel \\
	            \hline
	            $1/2$ & $(2, 1)$ & $20$  & bosonic, even parity  & (b) \\
	            $1/2$ & $(3, 1)$ & $40$  & fermionic, even parity & (e) \\
	            $1/3$ & $(4, 2)$ & $45$  & bosonic, odd parity   & (c) \\
	            $2/3$ & $(2, 1)$ & $45$  & fermionic, odd parity  & (f) \\
	            \end{tabular}
	        \end{ruledtabular}
    \end{table}

\begin{figure}[htbp]
        \centering
        \includegraphics[width= \linewidth]{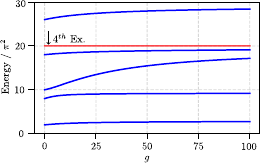}
        \caption{Energy vs. $g$ in the bosonic
        even parity sector at $c=1/2$. The $4^{\text{th}}$ excited state is dark.
        The $2^{\text{nd}}$ and $3^{\text{rd}}$ excited states approach
        the energy of the dark state asymptotically, giving rise to a triple
        degeneracy in the $g \to \infty$ limit.}
        \label{Fig:EvsG_Dark}
    \end{figure}

Fig.~\ref{Fig:EvsG_Dark} shows the energy as a function of the interaction strength $g$ at $c = 1/2$ in the 
bosonic sector with even parity.
As shown in this figure, the $4^{\text{th}}$ excited state is dark at this value of $c$:
its energy remains strictly insensitive to $g$, i.e., the interaction leaves this state unperturbed.
Note that in the limit $g\to\infty$,
the $2^{\text{nd}}$ and $3^{\text{rd}}$ excited states approach the energy of the dark state asymptotically, signaling a triple degeneracy.
The corresponding nodal structure of this wavefunction is shown in panel~(b) of Fig.~\ref{Fig:DarkStates}.

A transparent route to derive the necessary conditions for the emergence of dark states is to compare the analytical solutions
in the two limiting cases $g \to 0$ and $g \to \infty$.
A dark state, by definition, is an eigenstate in both these limits.
In the strong interaction limit, bosonic wavefunctions become fermionized~\cite{girardeau1960relationship}.
The problem can therefore be formulated as identifying the subset of fermionized eigenfunctions that are also eigenfunctions in the non-interacting limit.
Eigenfunctions that belong to this subset satisfy Eq.~(\ref{eq:dark_condition}) in both limits, signaling the presence of diagonal nodal lines at $x_2 = x_1 \pm c$, which give rise to dark states.

The $g \to 0$ fermionic eigenfunctions with quantum numbers $(N,M)$ in a unit box are
    \begin{multline}
        \Psi^{g \to 0}_{NM}(x_1,x_2) = 
        \cfrac{1}{\sqrt{2}}[\sin(N\pi x_1)\sin(M\pi x_2)\\
                -\sin(M\pi x_1)\sin(N\pi x_2)],
    \end{multline}
and satisfies Eq.~\eqref{eq:dark_condition} only if $N \pi$ and $M \pi$
coincide with quantized quasimomenta $k_n = \pi n/a$ of the
region outside spectrum in the $g \to \infty$ limit, that is,
    \begin{equation}
        N = \frac{n}{1-c} \in \mathbb{N},
        \qquad
        M = \frac{m}{1-c} \in \mathbb{N},
        \qquad n > m \geq 1.
        \label{eq:dark_condition_qn}
    \end{equation}
    The corresponding dark-state energies are
    \begin{equation}
        \varepsilon_{nm}
        = \frac{\pi^2(n^2+m^2)}{(1-c)^2}.
    \end{equation}
    Similar to the region outside in the $g \to \infty$ limit, energy levels with $n + m = 2k$ correspond to the bosonic, even-parity sector and the fermionic, odd-parity sector,
    while those with $n + m = 2k + 1$ correspond to the bosonic, odd-parity sector and the fermionic, even-parity sector.
    As discussed below, all dark states are only triply degenerate in the $g \to \infty$ limit.

    \begin{figure*}[t]
            \centering
            \includegraphics[width= \linewidth]{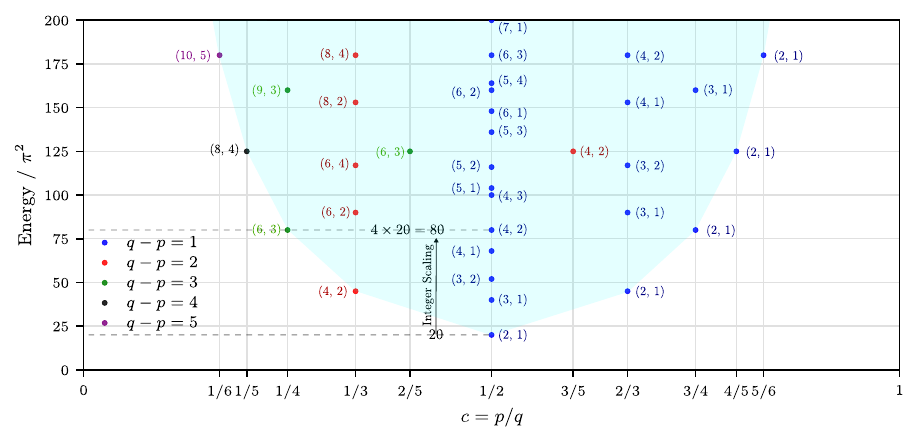}
            \caption{Dark state quantum numbers $(n,m)$ as a function of
            displacement $c = p/q$. For each point shown, an exact analytical solution exists.
	    Integer scaling generates an infinite tower at fixed $c$; only $(2,1)\to(4,2)$ at $c = 1/2$ is shown.}
            \label{Fig:DarkDistribution}
    \end{figure*}

The dark condition Eq.~(\ref{eq:dark_condition_qn}) requires $c = p / q$ to satisfy $(q-p)\mid n$ and $(q-p)\mid m$.
The densest set of dark states thus arises when $q-p=1$, i.e.,
    \begin{equation}
        c_k = \frac{k}{k+1}, \qquad k=1,2,3,\ldots,
        \label{eq:ck}
    \end{equation}
where all $(n,m)$ with $n>m\geq 1$ are admissible.

Dark states inherit the scaling properties of the non-interacting spectrum. Specifically, for a given dark state with quantum numbers $(n,m)$ at displacement
$c$, every state with quantum numbers $(jn, jm)$ for integer $j$ is also
a dark state at the same $c$, since
    \begin{equation}
        \frac{jn}{1-c},\quad \frac{jm}{1-c} \in \mathbb{N}
        \quad\Longleftrightarrow\quad
        \frac{n}{1-c},\quad \frac{m}{1-c} \in \mathbb{N}.
    \end{equation}
Each base dark state $(n_0,m_0)$ therefore generates an infinite tower of dark
states by integer scaling, with energies
    \begin{equation}
        \varepsilon_{j n_0,\, j m_0}
        = j^2\,\varepsilon_{n_0 m_0}.
    \end{equation}

The tower of dark states at a given $c$ spans a subspace of the
full Hilbert space of $\mathcal{H}_0 + V(g,c)$ in which states do not mix with the rest of the
spectrum at any interaction strength.
Within this subspace, the dynamics are governed by $\mathcal{H}_0$ alone,
making the tower an exactly solvable sector embedded in an otherwise non-integrable spectrum.
Therefore in the region outside, the wavefunction for the dark states is exactly those introduced in Eq.~(\ref{eq:triangle_eigenfunctions}) and shown in Fig.~\ref{Fig:Triangles}.

Fig.~\ref{Fig:DarkDistribution} displays all pairs $(n,m)$ which give rise to a dark state of the interaction,
as a function of $c$ up to $\varepsilon/\pi^2 = 200$, grouped by $q-p$ where $c = p/q$.
Note that as $c$ deviates from $1/2$, the corresponding dark states move higher in the spectrum.
For each value of $c$ in the spectrum, there exists a close rational value in the spectrum that corresponds, at sufficiently high energies, to a dark state.

We discussed in Sec.~\ref{subsec:inf_g} that in the $g \to \infty$ limit,
the energy spectrum in the region outside is two-fold degenerate.
A dark state, being simultaneously an eigenstate of the region outside and of the
free problem, sits at the same energy as this two-fold degenerate level in the region outside.
This promotes the degeneracy from two to three.
Thus, every dark state generates a triple degeneracy in the
$g\to\infty$ spectrum: Fig.~\ref{Fig:EvsG_Dark} for $c=1/2$ asymptotically signals this triple degeneracy.
This triple degeneracy at $c = 1/2$ is particularly visible in Fig.~\ref{Fig:Combined_EvsC_g_infty} (panel (a)) at $\varepsilon / \pi^2 = 20$.
Triple degeneracy is therefore a necessary condition for a dark state, but
not sufficient. The sufficient condition is Eq.~\eqref{eq:dark_condition_qn}: the triple degeneracy
must occur at an energy that is simultaneously a non-interacting energy $(N^2+M^2)\pi^2$
and an region outside energy $\pi^2(n^2+m^2)/(1-c)^2$ with $N=n/(1-c)$,
$M=m/(1-c)$ integers. All other triple degeneracies in the spectrum are regarded as accidental degeneracies.
Note that similar to the harmonic potential case, a small variation in $c$ induces a significant change in the spatial structure of the dark states.
In particular, if the $n^{\text{th}}$ excited state at a given $c$ is a dark state, then a small change in $c$ reduces the triple degeneracy to a single non-degenerate state, whose wavefunction becomes localized in the region inside, with its amplitude in the region outside being strongly suppressed.

\section{Conclusions}
\label{sec:Conclusion}
We studied the stationary properties of two particles trapped in a hard-wall box, interacting via a displaced contact interaction with displacement $c$.
In particular, we studied three exactly solvable special cases.
In the non-interacting limit, the system is exactly solvable by separation of variables, 
and the full energy spectrum is analytically accessible from the single-particle solutions. 
For the contact interaction ($c = 0$), the system is Bethe-ansatz solvable.
In this case, we derived the exact bosonic spectrum from coupled transcendental equations and used it to benchmark our exact diagonalization (ED) approach.
We also discussed that the fermionic spectrum in this case remains unperturbed by the interaction.
In the strongly interacting limit ($g \to \infty$) at arbitrary $c$,
the problem is equivalent to quantum billiards in polygons:
the region outside corresponds to an integrable quantum billiard in a $(\pi/4,\,\pi/4,\,\pi/2)$ triangle, which is exactly solvable,
while the region inside corresponds to a pseudointegrable billiard and must be solved numerically via ED. 
In the latter case, the spectrum for the region outside is doubly degenerate,
and all crossings between outer- and inner-region levels are exact, giving rise to four distinct 
regimes. In the \emph{exclusion} regime~(E), only doubly degenerate outer-region
levels are present, and both particles reside in the region outside. In the
\emph{truncation} regime~(T), only nondegenerate inner-region levels exist, and
the particles are localized into the region inside. Between these two extremes,
two \emph{crossover} regimes~(C) exist in which outer- and inner-region levels
coexist: in the first, outer-region levels coexist with only the bosonic
inner-region states; in the second, they coexist with both bosonic and fermionic
inner-region states. In both crossover regimes, degenerate and nondegenerate
eigenstates are simultaneously present. 

We also analyzed the dark states in our model:
Dark states exist when $c$ is rational and the quantum numbers 
$(N, M)$ are integer multiples of $1/(1-c)$; each primitive dark state generates 
an infinite tower of dark states with energies scaling as $j^2$, constituting an 
exactly solvable sector embedded in an otherwise non-integrable spectrum. In the 
$g \to \infty$ limit, dark states manifest as triple degeneracies and reside 
in the crossover regimes~(C), similar to the behavior found in the harmonic 
potential case~\cite{bougasImpactDarkStates2024}.

Dark states and partial solvability appear to be a generic feature of zero-range potentials. 
Generally, a potential given by one or more N-dimensional delta functions defines an intersection of manifolds with co-dimension N in configuration space.
When the nodal structures of wavefunctions without the potential align with this intersection,
then the state is dark. 
For the two-body contact interaction in one dimension, the nodal structure of fermionic states is forced to coincidence, 
with the potential by exchange antisymmetry. 
Here, we explore a more generic manifestation, where the nodal structure is not guaranteed by a symmetry. It is therefore less robust to deviations in trap shape,
but it also reveals novel possibilities for engineering systems with number theory built into their spectrum. 

\appendix
\section{Analytical evaluation of the potential matrix elements}
\label{app:PotentialMatrix}
In this Appendix, we present an analytical evaluation of the potential matrix elements $\mathcal{V}_{nm,n'm'}$.
The integration limits in Eq.~(\ref{eq:V_elements}) are determined by the boundary conditions $0 \le x_1, x_2 \le 1$:
for the term $\delta(x_2-x_1+c)$, the condition $0 \le x_2 = x_1-c \le 1$ requires $c \le x_1 \le 1$.
Similarly, for the term $\delta(x_2 - x_1 - c)$, $0 \le x_2 = x_1+c \le 1$ implies $0 \le x_1 \le 1-c$.
The first integral contributes only in the domain $x_2 \leq x_1$, and the second contributes only when $x_2 \ge x_1$.

Both integrands in Eq.~(\ref{eq:V_elements}) contain terms of the form
\begin{equation}
    4 \sin(n\pi x_1)\sin(m\pi(x_1\pm c))\sin(n'\pi x_1)\sin(m'\pi(x_1\pm c)).
    \label{eq:sin_terms}
\end{equation}
Using the relation
    \begin{equation}
        \sin A\,\sin B=\tfrac12\big[\cos(A-B)-\cos(A+B)\big],
    \end{equation}
we obtain
    \begin{multline}
       4 \sin(n\pi x_1)\sin(m\pi(x_1\pm c))\sin(n'\pi x_1)\sin(m'\pi(x_1\pm c)) \\
        \qquad = \cos \big[(n-n')\pi x_1\big]\cos \big[(m-m')\pi(x_1\pm c)\big] \\
        \qquad -\cos \big[(n-n')\pi x_1\big]\cos \big[(m+m')\pi(x_1\pm c)\big]\\
        \qquad  -\cos \big[(n+n')\pi x_1\big]\cos \big[(m-m')\pi(x_1\pm c)\big]\\
        \qquad +\cos \big[(n+n')\pi x_1\big]\cos \big[(m+m')\pi(x_1\pm c)\big].
        \label{eq:4sin_expanded}
    \end{multline}
We define
    \begin{equation}
        \mathcal T_{p\!,\,q}^{(\pm)}\;:=\;
        \int_{\mathcal D_\pm}\cos(p\pi x_1)\,\cos\!\big(q\pi (x_1\pm c)\big)\,dx_1,
        \label{eq:Tdef}
    \end{equation}
where \(p,q\in\{n\pm n',m\pm m'\}\) and the
integration domain \(\mathcal D_\pm\):
    \begin{equation*}
        \mathcal D_{-}=[c,1],\qquad \mathcal D_{+}=[0,1-c].
    \end{equation*}
$\mathcal{T}_{p,q}^{(\pm)}$ give rise to five different cases depending on whether $p$ and $q$ are zero.
The closed forms for each case are given below:

\noindent\textbf{Case (i): $p\neq q \neq 0$}
    \begin{multline}
        \mathcal{T}_{p,q}^{(+)} =  
        \frac{2q \sin(c q \pi) - (p + q)\sin\big[((c - 1)p + q)\pi\big]}{2(p - q)(p + q)\pi} \\
        \quad + \frac{(p - q)\sin\big[((1 - c)p + q)\pi\big]}{2(p - q)(p + q)\pi}
        \label{eq:T_general_plus}
    \end{multline}

    \begin{multline}
        \mathcal{T}_{p,q}^{(-)} =  
        \frac{-2p \sin(c p \pi) + (p + q)\sin\big[(p + (c - 1)q)\pi\big]}{2(p - q)(p + q)\pi} \\
        \quad + \frac{(p - q)\sin\big[(p + (1 - c) q)\pi\big]}{2(p - q)(p + q)\pi}
        \label{eq:T_general_minus}
    \end{multline}

\noindent\textbf{Case (ii): $p = \pm q \neq 0$}
    \begin{equation}
        \mathcal{T}_{\pm q,q}^{(\pm)} =-\frac{\sin [(c-2) q \pi]+\sin (c q \pi)+2 \pi  (c-1) q \cos (c q \pi)}{4 q \pi}
        \label{eq:T_equal_minus}
    \end{equation}

\noindent\textbf{Case (iii): $p = 0$, $q \neq 0$}
    \begin{equation}
        \mathcal{T}_{0,q}^{(+)} = \frac{\sin(q\pi)-\sin(cq\pi)}{q\pi}
        \label{eq:T_p0_plus}
    \end{equation}

    \begin{equation}
        \mathcal{T}_{0,q}^{(-)} = \frac{\sin\big[(1-c)q\pi\big]}{q\pi}
        \label{eq:T_p0_minus}
    \end{equation}

\noindent\textbf{Case (iv): $p \neq 0$, $q = 0$}
    \begin{equation}
        \mathcal{T}_{p,0}^{(+)} = \frac{\sin\big[(1-c)p\pi\big]}{p\pi}
        \label{eq:T_q0_plus}
    \end{equation}

    \begin{equation}
        \mathcal{T}_{p,0}^{(-)} = \frac{\sin(p\pi)-\sin(cp\pi)}{p\pi}
        \label{eq:T_q0_minus}
    \end{equation}

\noindent\textbf{Case (v): $p = q = 0$}
        \begin{equation}
        \mathcal{T}_{0,0}^{(\pm)} = 1-c
        \label{eq:T_00}
    \end{equation}

The integral of (\ref{eq:sin_terms}) can now be expressed as:
    \begin{multline}
        \mathcal{S}_{[n n' ,m m']}^{(\pm)} = \mathcal{T}_{n-n',m-m'}^{(\pm)}-\mathcal{T}_{n-n',m+m'}^{(\pm)} \\ 
 \quad -\mathcal{T}_{n+n',m-m'}^{(\pm)}+\mathcal{T}_{n+n',m+m'}^{(\pm)}
        \label{eq:S_element}
    \end{multline}
The complete matrix element, including normalization of $\psi_{nm}$ is given by
    \begin{multline}
        \mathcal{V}_{nm,n'm'} = 
        \frac{g}{2 \sqrt{(1+\delta_{nm})(1+\delta_{n'm'})}} \times \\
        \quad\Big[\mathcal{S}_{[n n', m m']}^{(+)} +
        \mathcal{S}_{[n n', m m']}^{(-)} 
        \pm \mathcal{S}_{[m n', m' n]}^{(-)} 
        \pm \mathcal{S}_{[m n', m' n]}^{(+)} \\
        \qquad\pm \mathcal{S}_{[m' n, m n']}^{(-)} 
        \pm \mathcal{S}_{[m'n, m n']}^{(+)} 
        + \mathcal{S}_{[m m', n n']}^{(-)} 
        + \mathcal{S}_{[m m', n n']}^{(+)}
        \Big].
        \label{eq:V_final}
    \end{multline}
The upper signs between the $\mathcal{S}$ terms correspond to bosonic sectors, while the lower signs correspond to fermionic sectors.

\section{Bethe-ansatz derivation for c=0}
\label{app:bethe}
We outline the derivation of Eqs.~\eqref{eq:bethe_main} in dimensionless units with $L=1$.
The Bethe ansatz wavefunction for the bosonic sector is given by \cite{Batchelor_2005},
    \begin{multline}
        \Psi(x_1,x_2) = \\ \sum_{\substack{\epsilon_1,\epsilon_2=\pm 1\\\sigma\in S_2}}
        \epsilon_1\epsilon_2\,
        A(\epsilon_1 k_{\sigma(1)},\epsilon_2 k_{\sigma(2)}) 
        e^{i(\epsilon_1 k_{\sigma(1)}x_1+\epsilon_2 k_{\sigma(2)}x_2)}.
        \label{eq:bethe_wf}
    \end{multline}
$A(\epsilon_1 k_{\sigma(1)},\epsilon_2 k_{\sigma(2)})$ encodes the two body scattering matrix.
For $x_1 \neq x_2$, $\mathcal{H}\Psi(x_1,x_2) = (k_1^2 + k_2^2)\Psi(x_1,x_2)$.
Integrating the Schr\"{o}dinger equation across $x_1=x_2$ yields the first-derivative jump condition. For bosonic wavefunctions:
    \begin{equation}
        (\partial_{x_2} - \partial_{x_1})
        \Psi\big|_{x_2=x_1} = g\,\Psi(x_1,x_2).
        \label{eq:jump_boson_app}
    \end{equation}
For fermionic wavefunctions, antisymmetry forces $\Psi|_{x_1=x_2}=0$, so the
right hand side of Eq.~(\ref{eq:jump_boson_app}) vanishes; the fermionic sector
is therefore dark to the interaction for all $g$, with energies
    \begin{equation}
        \varepsilon_{nm}^{(f)} = \pi^2(n^2 + m^2), \qquad n > m\geq 1.
    \end{equation}
The fermionic wavefunction is simply the non-interacting basis function listed in Table~\ref{tab:symmetries} for $\sigma = -1$.

For the bosonic sector, imposing hard-wall boundary conditions at $x_j=0$ and $x_j=1$ combining with Eq.~\eqref{eq:jump_boson_app} gives~\cite{Batchelor_2005}
    \begin{subequations}
    \label{eq:bethe_exp}
        \begin{align}
            e^{2ik_1} &= \frac{(k_1-k_2+ig)(k_1+k_2+ig)}{(k_1-k_2-ig)(k_1+k_2-ig)}, \\
            e^{2ik_2} &= \frac{(k_2-k_1+ig)(k_2+k_1+ig)}{(k_2-k_1-ig)(k_2+k_1-ig)}.
        \end{align}
    \end{subequations}
Taking the principal-value logarithm, introducing $K=k_1+k_2$ and $\Delta=k_1-k_2$, and using the identity
    \begin{equation}
        \frac{i}{2}\ln\frac{q-ig}{q+ig} = \arctan\!\left(\frac{g}{q}\right),
        \label{eq:log_arctan}
    \end{equation}
decouples the equations into the pair~\eqref{eq:bethe_main} in the main text, with eigenenergy
    \begin{equation}
        \varepsilon^{(b)}_{n m} = \frac{1}{2}\!\left(\Delta_{nm}^2+K_{nm}^2\right).
    \end{equation}

\begin{acknowledgments}
The authors would like to acknowledge useful discussions with Maxim Olshanii on the subject of integrability and solvability.
\end{acknowledgments}

\bibliography{DecenteredDeltas.bib}
\end{document}